\documentstyle [12pt]{article}

\begin{document}

\vspace{2mm}

\begin{flushright}
Preprint MRI-PHY/8/95 \\

hepth/9505151, May 1995
\end{flushright}

\vspace{2ex}

\begin{center}

{\large \bf Implications of Some Static Spherically

\vspace{2ex}

            Symmetric Graviton-Dilaton Solutions in

\vspace{2ex}

            Brans-Dicke and Low Energy String Theory\footnote{Invited
Talk presented at the XI DAE Symposium held during
21 - 28 December 1994 in Shantiniketan, India.}  } \\

\vspace{6mm}
{\large S. Kalyana Rama}
\vspace{3mm}

Mehta Research Institute, 10 Kasturba Gandhi Marg,

Allahabad 211 002, India.

\vspace{1ex}
email: krama@mri.ernet.in   \\
\end{center}

\vspace{4mm}

\begin{quote}
ABSTRACT.  Analysing the static, spherically symmetric \\
graviton-dilaton
solutions in low energy string and Brans-Dicke theory, we find the
following. For a charge neutral point star, these theories cannot predict
non trivial PPN parameters, $\beta$ and $\gamma$, without introducing
naked singularities. We then couple a cosmological constant $\Lambda$ as
in low energy string theory. We find that only in low energy string
theory, a non zero $\Lambda$ leads to a curvature singularity, which is
much worse than a naked singularity. Requiring its absence upto
a distance $r_*$ implies a bound $| \Lambda | < 10^{- 102} (\frac{r_*}
{{\rm pc}})^{- 2}$ in natural units. If $r_* \simeq 1 {\rm Mpc}$ then
$| \Lambda | < 10^{- 114}$ and, if $r_* \simeq 10^{28} {\rm cm}$ then
$| \Lambda | < 10^{- 122}$ in natural units.

\vspace{4ex}


\end{quote}

\newpage

\vspace{4ex}

{\bf 1.}
We study the static spherically symmetric solutions for
Brans-Dicke (BD) and low energy string
theory, including only the graviton and the dilaton field. They describe
the gravitational field of a charge neutral point star, in these
theories. Calculating the parametrised post Newtonian (PPN) parameters
$\beta$ and $\gamma$ \cite{will},
we find that all the acceptable solutions predict
$\beta = \gamma = 1$, the same as in Einstein's theory. There are more
general static spherically symmetric solutions \cite{b}-\cite{tcd6}
predicting $\beta = 1, \;
\gamma = 1 + \epsilon$, but they always have naked curvature
singularities proportional to $\epsilon^2$ and, hence, are unacceptable.

These general
solutions can be better understood by coupling the electromagnetic
field \cite{gm,ghs}. They lead to non trivial PPN parameters for
a point star of charge $Q$ . In these solutions there is
an inner and an outer horizon. The curvature scalar is singular
at the inner horizon, but this singularity is
hidden behind the outer horizon. A charge neutral star can
then be obtained in two ways: in one, corresponding to the Schwarzschild
solution, the PPN parameters are trivial and
there is no naked singularity, while in the other, the PPN parameters
are non trivial but there is a naked singularity.

Therefore neither BD nor low energy string theory can predict non
trivial values for
PPN parameters $\beta$ and $\gamma$, for a charge neutral star,
without introducing naked singularities.

We also couple a cosmological constant $\Lambda$, in a way analogous
to the coupling of a tree level cosmological constant in low energy
string theory \cite{tdgang}. The static spherically symmetric solutions
here describe the gravitational field of point stars, and it is
reasonable to expect them to be valid upto a distance $r_*$, of ${\cal O}
({\rm pc})$, even when the real universe is not static but expanding.

{}From an analysis of these solutions, we find \cite{kcc} that for low
energy string theory, a non zero $\Lambda$ leads to a curvature
singularity which is physically unacceptable. Requiring their absence
imposes a bound $| \Lambda | < 10^{- 102} (\frac{r_*}{{\rm pc}})^{- 2}$
in natural units. Thus if $r_* \simeq 1 {\rm Mpc}$ then
$| \Lambda | < 10^{- 114}$, and if $r_*$ extends all the way upto the edge
of the universe ($10^{28} {\rm cm}$) then $| \Lambda | < 10^{- 122}$
in natural units. For more details see \cite{kppn}. For static
solutions in other contexts, see \cite{wilt}

\vspace{4ex}

{\bf 2.} Consider the following action for
graviton $(\tilde{g}_{\mu \nu})$ and  dilaton $(\phi)$ fields,
\begin{equation}\label{starget}
S = - \frac{1}{16 \pi}
\int d^4 x \sqrt{\tilde{g}} \, e^{\phi} \,
( \tilde{R} - \tilde{a} (\tilde{\nabla} \phi)^2 + \Lambda )
\end{equation}
where $\Lambda$ is the cosmological constant coupled to $\phi$, as
in string theory, and $R_{\mu \nu \lambda \tau} =
\frac{\partial^2 g_{\mu \lambda}} {\partial x^{\nu} \partial x^{\tau}}
+ \cdots$. For low energy string theory $\tilde{a} = 1$, and for BD
theory $\tilde{a} = - \omega$, the BD parameter.

It is easier to solve the equations of motion if one makes the
transformation $\tilde{g}_{\mu \nu} = e^{- \phi} g_{\mu \nu}$.
The physical curvature scalars is given by
\begin{equation}\label{rstring}
\tilde{R} = e^{\phi}
( R - 3 \nabla^2 \phi + \frac{3}{2} (\nabla \phi)^2 )
\end{equation}
where $R$ is the curvature scalar obtained using $g_{\mu \nu}$.
The equations of motion become
\begin{eqnarray}\label{beta}
2 R_{\mu \nu} + a \nabla_{\mu} \phi \nabla_{\nu} \phi
+ g_{\mu \nu} \Lambda e^{- \phi}  & = & 0 \nonumber \\
a \nabla^2 \phi + \Lambda e^{- \phi} & = & 0 \; ,
\end{eqnarray}
where $a \equiv 3 - 2 \tilde{a}$. We study the static, spherically
symmetric solutions to equations (\ref{beta}).

If $\Lambda = 0$, we take the metric to be
$d s^2 = - f d t^2 + f^{- 1} d \rho^2 + r^2 d \Omega^2$, where the fields
$f, \; r$, and $\phi$ depend only on $\rho$. If $\Lambda \ne 0$, we take
the metric to be $d s^2 = - f d t^2 + \frac{G}{f} d r^2 + r^2
d \Omega^2$, where the fields $f, \; G$, and $\phi$ depend only on $r$.
In these expressions, $d \Omega^2$ is the line element on an unit sphere.
In the first case, the equations (\ref{beta}) become
\begin{eqnarray}\label{rf}
\frac{(f r^2)''}{2} - 1 & = & ( f' r^2 )' \nonumber \\
= a ( \phi' f r^2 )' - \Lambda_{\phi} r^2 e^{- \phi}
& = & - \Lambda r^2 e^{- \phi} \nonumber \\
4 r'' + a r \phi'^2 & = & 0
\end{eqnarray}
where $'$ denotes $\rho$-derivatives. In the second case, they become
\begin{eqnarray}\label{gf}
\frac{(f r^2)''}{2} - \frac{(f r^2)' G'}{4 G} - G
& = & ( f' r^2 )' - \frac{G' f' r^2}{2 G}
\nonumber \\
= a ( \phi' f r^2 )' - \frac{a \phi' G' f r^2}{2 G}
- \Lambda_{\phi} G r^2 e^{- \phi}
& = & - \Lambda G r^2 e^{- \phi} \nonumber \\
2 G' - a r G \phi'^2 & = & 0
\end{eqnarray}
where $'$ denotes $r$-derivatives, and the physical
curvature scalar $\tilde{R}$ is given by
\begin{equation}\label{r}
\tilde{R} = \frac{(3 - a) f \phi'^2 e^{\phi}}{2 G}
+ \frac{(3 - 2 a) \Lambda}{a}  \; .
\end{equation}
By writing the physical metric $\tilde{g}_{\mu \nu}$ in isotropic form,
and expanding its $tt$ and $rr$ components asymptotically, one obtains
the PPN parameters. For details, see \cite{will}.

\vspace{4ex}

{\bf 3.} When $\Lambda = 0$, the most general solutions to equations
(\ref{rf}) are given by \cite{b,tcd6}
\begin{equation}\label{bdsoln}
f = \left( 1 - \frac{\rho_0}{\rho} \right)^{\frac{1 - k^2}{1 + k^2}}
\; , \; \;
r^2  =
\rho^2 \left( 1 - \frac{\rho_0}{\rho} \right)^{\frac{2 k^2}{1 + k^2}}
\; , \; \;
e^{\phi - \phi_0} =
\left( 1 - \frac{\rho_0}{\rho} \right)^{\frac{2 l}{1 + k^2}}
\end{equation}
where $k$ is a parameter and $l \equiv \frac{k}{\sqrt{a}}$.
Writing the physical metric $\tilde{g}_{\mu \nu}$
in the isotropic gauge, $d \tilde{s}^2 \equiv - \tilde{f} d t^2
+ \tilde{F} (d \rho^2 + \rho^2 d \Omega^2)$, where $r$ and $\rho$ are
related by
\begin{equation}\label{hrho}
\rho = h \left( 1 + \frac{\rho_0}{4 h} \right)^2 \; ,
\end{equation}
the physical mass $M$ and the PPN parameters $\beta$ and $\gamma$ are
given, after a straightforward calculation, by
\begin{equation}\label{mass}
2 M = \frac{1 - k^2 - 2 l}{1 + k^2} \rho_0  \; , \; \;
\beta = 1 \; , \; \;
\gamma = 1 + \frac{2 l \rho_0}{(1 + k^2) M}  \; .
\end{equation}
The parameter $\beta$ is trivial while $\gamma$ is non trivial if
$l \rho_0 \ne 0$. The physical curvature scalar $\tilde{R}$ is given by
\begin{equation}\label{rtilde1}
\tilde{R} = \frac{\tilde{a} M^2 (\gamma - 1)^2 e^{\phi_0}}{\rho^4} \;
\left(1 - \frac{\rho_0}{\rho}\right)^{- \frac{1 + 3 k^2 - 2 l}{1 + k^2}}
\; .
\end{equation}
In the above equations $\rho_0$ is positive,
so that one obtains the standard Schwarzschild solution when $k = 0$.
Also the physical mass $M$, given by (\ref{mass}), must
be positive which then implies that $1 - k^2 - 2 l > 0$. Hence,
the metric component $\tilde{g}_{tt}$ in the physical frame vanishes at
$\rho = \rho_0$. The above condition on $k$ also implies that
$1 + 3 k^2 - 2 l >0$. Hence, the curvature scalar $\tilde{R}$
in (\ref{rtilde1}) becomes singular there, unless
$\gamma = 1$, {\em i.e.}\ unless the PPN parameters are trivial.
This singularity is naked, as will be shown presently.

The experimentally observed range of the PPN parameter $\gamma$ is
$\gamma = 1 \pm .002$. Requiring $|\gamma - 1| < \epsilon < .002$,
and taking into account the constraint $1 - k^2 - 2 l > 0$,
restricts $k$ to be
\begin{equation}\label{k2}
|k| < \frac{\epsilon \sqrt{a}}{2 (1 + \epsilon)} \; .
\end{equation}

Now we will discuss the nature of the singularity at $\rho = \rho_0$.

\noindent 1. As can be seen from equation (\ref{rtilde1}), the
curvature scalar is singular at $\rho = \rho_0$; hence, this singularity
is not a coordinate artifact and cannot be removed by any coordinate
transformation.

\noindent 2. The metric on the surface $\rho = \rho_0$ has the signature
$0+++$, and hence, this surface is null and the singularity is a null
one.

\noindent 3. Consider an outgoing radial null geodesic, which
describes an outgoing photon. Since $d \tilde{s}^2 = 0$
for such a geodesic, its equation is given by
\[
\frac{d t}{d \rho} =
\left( 1 - \frac{\rho_0}{\rho} \right)^{\frac{k^2 - 1}{k^2 + 1}} \; ,
\]
where $t$ is the external time, which gives $t = \rho_* + const$,
where $\rho_*$, the analog of the `tortoise coordinate', is defined by
\[
\rho_* = \int d \rho
\left( 1 - \frac{\rho_0}{\rho} \right)^{\frac{k^2 - 1}{k^2 + 1}} \; .
\]
For $k \ne 0$, it is easy to show that $\rho_* (\rho)$ is finite.
The outgoing radial null geodesic equation given above then implies
that a radially outgoing photon starting from $\rho_i \; ( \ge \rho_0 )$
at external time $t_i$ will reach an outside observer at
$\rho_f \; (\rho_i < \rho_f < \infty)$ at a finite external time
$t_f$ given by $t_f - t_i = \rho_*(\rho_f) - \rho_*(\rho_i)$. Hence,
it follows that a photon can travel from
arbitrarily close to the singularity to an outside observer
within a finite external time interval and, therefore, the singularity
at $\rho = \rho_0$ is naked.

For these reasons, the singularity at $\rho = \rho_0$ is naked and
physically  unacceptable. For recent detailed discussions on
naked singularities and their various general aspects see \cite{psj}).

One can gain more insight into the solution (\ref{bdsoln}) by coupling
a $U(1)$ gauge field $A_{\mu}$, as in \cite{gm,ghs}, with field
strength $F_{\mu \nu}$. The general solution is then given by
\begin{eqnarray}
f & = & \left( 1 - \frac{\rho_1}{\rho} \right)
\left( 1 - \frac{\rho_0}{\rho} \right)^{\frac{1 - k^2}{1 + k^2}}
\; , \; \;
r^2 = \rho^2
\left( 1 - \frac{\rho_0}{\rho} \right)^{\frac{2 k^2}{1 + k^2}}
\nonumber \\
e^{\phi - \phi_0} & = &
\left( 1 - \frac{\rho_0}{\rho} \right)^{\frac{2 l}{1 + k^2}} \; , \; \;
F_{t \rho} = \frac{Q}{\rho^2}
\end{eqnarray}
where $l = \frac{k}{\sqrt{a}}$ and the remaining components of
$F_{\mu \nu}$ are zero. Writing the physical metric $\tilde{g}_{\mu \nu}$
in the isotropic gauge as before, the physical parameters
$M, \; Q, \; \beta$, and $\gamma$ are given by
\begin{eqnarray*}
2 M & = & \rho_1 + \frac{1 - k^2 - 2 l}{1 + k^2} \rho_0 \; , \; \;
Q^2 = \frac{\rho_1 \rho_0}{1 + k^2} \\
\beta & = & 1 + \frac{(1 - l) Q^2}{2 M^2} \; , \; \;
\gamma = 1 + \frac{2 l \rho_0}{(1 + k^2) M}  \; .
\end{eqnarray*}
The parameter $\beta$ is non trivial if the charge $Q \ne 0$ while
$\gamma$ is non trivial if $l \rho_0 \ne 0$.

The curvature scalar $\tilde{R}$ in the physical frame is given by
\begin{equation}\label{rtilde2}
\tilde{R} = \frac{\tilde{a} M^2 (\gamma - 1)^2 e^{\phi_0}}{\rho^4} \;
\left(1 - \frac{\rho_1}{\rho}\right) \;
\left(1 - \frac{\rho_0}{\rho}\right)^{- \frac{1 + 3 k^2 - 2 l}{1 + k^2}}
\; .
\end{equation}
The metric component $\tilde{g}_{tt}$ in the physical frame vanishes at
$\rho = \rho_1$ and $\rho = \rho_0$. The curvature scalar
$\tilde{R}$ is regular at $\rho = \rho_1$ but,
since $1 + 3 k^2 - 2 l > 0$ for $a \ge 1$, it is singular
at $\rho = \rho_0$ unless $\gamma = 1$. This singularity is
hidden behind the horizon at $\rho_1$
if $\rho_1 > \rho_0$, and naked otherwise for the same reasons as given
following equation (\ref{rtilde1}).

Now, a charge neutral solution {\em i.e.}\ $Q = 0$, can be obtained by
setting either $\rho_0 = 0$ or $\rho_1 = 0$. The first case corresponds
to the Schwarzschild solution while the second one, to the solution
(\ref{bdsoln}) for which $\gamma$ is non trivial.

Thus, it follows that in BD or low energy string theory, a non trivial
value for $\gamma$ for a charge neutral point star implies the existence
of a naked singularity.
Conversely, in these theories, the absence of naked
singularities necessarily implies that the PPN parameters $\beta$
and $\gamma$ for a charge neutral point star are trivial.

\vspace{4ex}

{\bf 4.}
In the presence of both the dilaton $\phi$, and a non zero cosmological
constant $\Lambda \ne 0$, the solution to equations (\ref{gf}) is not
known in an explicit form. Here we study this solution and its
implications. Any solution, required to reduce to the Schwarzschild one
when $\Lambda = 0$, has the following general features:

(i)   The dilaton field $\phi$ cannot be a constant. The only exception
is when $\Lambda = \lambda e^{\phi}$, which corresponds to Einstein
theory with a cosmological constant $\lambda$ and a free field $\phi$.

(ii) In equations (\ref{gf}),  $\ln G$ and, hence $G$, strictly
increases since $a \ge 1$ and consequently $(\ln G)' > 0$.

(iii) Consider the following polynomial
ansatz for the fields as $r \to \infty$.
\begin{equation}
f = A r^k + \cdots \; , \; \;
G = B r^l + \cdots \; , \; \;
e^{- \phi} = e^{- \phi_0}  r^m + \cdots
\end{equation}
where $\cdots$ denote subleading terms in the limit $r \to \infty$.
Substituting these expressions into equations (\ref{gf})
gives, to the leading order, $2 l = a m^2$ and
\begin{eqnarray}\label{asym}
\frac{(k + 2)}{2} (k + 1 - \frac{l}{2}) A r^k - B r^l
& = & k (k + 1 - \frac{l}{2}) A r^k  \nonumber \\
= - a m (k + 1 - \frac{l}{2}) A r^k
& = & - B \Lambda e^{- \phi_0} B r^{l + m + 2}  \; .
\end{eqnarray}
The solution turns out to be
$(k, l, m) = (2 a, 2 a, - 2)$ or $(2, \frac{2}{a}, - \frac{2}{a})$, and
\[
k (k + 1 - \frac{l}{2}) A = - \Lambda e^{- \phi_0} B \; , \; \;
\left( (m + \frac{2}{a}) \Lambda e^{- \phi_0}
+ \frac{4}{r^{m + 2}} \right) B  = 0 \; .
\]

It is easy to see that if $a > 1$, as in BD theory, then there is always
a non trivial asymptotic solution with non zero $A$ and $B$. Also,
the physical curvature scalar $\tilde{R}$ for this solution is finite as
$r \to \infty$. Therefore it is plausible that a full solution can be
constructed with this asymptotic behaviour, which reduces to the
Schwarzschild solution when $\Lambda = 0$.

However, if $a = 1$ as in low energy string theory, then the above
equations are consistent only if $A = B = 0$. Hence, in this case,
equations (\ref{gf}) do not admit a non trivial solution where the
fields are polynomials in $r$ as $r \to \infty$. A similar analysis will
rule out asymptotic polynomial-logarithmic solutions, {\em i.e.}\ where
the fields behave as $r^m (\ln^n r) (\ln^p\ln r) + \cdots$ as
$r \to \infty$. Thus, for low energy string theory, the solutions cannot
have such asymptotic behaviour.

{}From now on, let $a = 1$. For small $r$, the solutions to (\ref{gf})
are given by
\begin{eqnarray}\label{fp}
f & = & 1 - \frac{r_0}{r} - \frac{\Lambda r^2}{6}
- \frac{\Lambda^2 r^4}{120} u_2 + \cdots \nonumber \\
G & = & 1 + \frac{\Lambda^2 r^4}{72} v_2 + \cdots \nonumber \\
\phi & = & \phi_0 - \frac{\Lambda r^2}{6} (1 + \frac{2 r_0}{r}
+ \frac{2 r_0^2}{r^2} \ln (r - r_0)) - \frac{\Lambda^2 r^4}{45} w_2
+ \cdots
\end{eqnarray}
where $\phi_0$ is a constant which can be set to zero,
and $u_i, \; v_i, \; w_i$ are functions of $\frac{r_0}{r}$
and $\ln r$ which tend to $1$ in the limit $\frac{r_0}{r} \ll 1$.
Evaluating further higher order terms will
not illuminate the general features of the solution. Also, the
series will typically have a finite radius of convergence beyond which
it is meaningless. Hence we follow a different approach.

It turns out that one can understand the general features
of the solutions using only the equations (\ref{gf}),
the behaviour of the fields for small $r$, and their
non polynomial-logarithmic behaviour as $r \to \infty$.

Note that $G = 1$ for Schwarzschild solution. Let $G$ has no pole
at any finite $r$. Then the requirement
that any solution to (\ref{gf}) reduce to the Schwarzschild one when
$\Lambda = 0$, combined with the fact that $G$ is a non decreasing
function, implies that $G (\infty)$ and, hence, $B$ must be non zero.
Then the above analysis, which excludes polynomial behaviour for the
fields with non trivial coefficients, implies in particular, that
the fields cannot be constant, including zero, as $r \to \infty$.

Consider first the case where $r_0 = 0$. From equations (\ref{gf}),
one then gets gives $e^{\phi} = |f|$. Also (see (\ref{fp})), $f$
has a local maximum (minimum) at the origin if $\Lambda$ is positive
(negative). Away from the origin, the function $f$ can
(A) have no pole at any finite $r$ and go to either $\infty$
or a constant as $r \to \infty$, or (B) have a pole at a finite $r = r_p$
(its behaviour for $r > r_p$ will not be necessary for our purposes).
We will also consider the case where (C) $f$ has a zero at $r = r_H$.

Case A: The function $f$, and hence $G$, has no pole at finite $r$. From
the above analysis, it follows that $f (\infty)$ cannot be a constant.
Hence, $f (\infty) \to \infty$, which also follows from the asymptotic
non polynomial-logarithmic behaviour of $f$.

Whether these singularities are genuine or only coordinate
artifacts can be decided by evaluating the curvature scalar, $\tilde{R}$,
or equivalently $R_1 \equiv \frac{f \phi'^2 e^{\phi}}{G}$
which can be shown, using (\ref{gf}), to obey the equation
\begin{equation}\label{r10}
R'_1 + \frac{4 R_1}{r} = - 2 \Lambda \frac{f'}{f}   \;   .
\end{equation}
Now, $R_1 (\infty)$ cannot be a constant. For, if it were, then one gets
$f (\infty) \to r^{- \frac{2 R_1 (\infty)}{\Lambda}}$, a polynomial
behaviour for $f$ as $r \to \infty$, which is ruled out.
Equation (\ref{r10}) can be solved to give
\[
R_1 = - \frac{2 \Lambda}{r^4} \; \int dr \frac{r^4 f'}{f} \; .
\]
{}From this it follows, as $r \to \infty$, that $\frac{f'}{f} > \frac{k}{r}$
for any constant $k$ (otherwise $R_1 (\infty) \to constant$).
This implies that $R_1 (\infty) \to \infty$.

Case B: The function $f$ has a pole at a finite $r = r_p < \infty$.
Then, from equation (\ref{r10}) it follows, near $r = r_p$,
that
\[
R_1 (r_p) = - 2 \Lambda \ln f (r_p) + {\cal O} (r - r_p)
\; \; \; \;  \to \; \; \pm \infty \; .
\]

Case C: The function $f$ has a zero at $r = r_H$. Then,
from equation (\ref{r10}) it follows, near $r = r_H$, that
\[
R_1 (r_H) = - 2 \Lambda \ln f (r_H) + {\cal O} (r - r_H)
\; \; \; \;  \to \; \; \pm \infty \; .
\]

Thus we see that $R_1$, and hence, the curvature scalar $\tilde{R}$
in the string frame, always diverges at one or more points
$r \equiv r_s = r_p, \; r_H, \; \infty$,
in low energy string theory when the cosmological
constant $\Lambda \ne 0$. These singularities, which will persist
even when $r_0 \ne 0$ as argued below, are naked.
In fact, they are much worse, as they are created by any object,
no matter how small its mass is. Thus at any
point of the string target space, there will be a singularity produced
by an object located at a distance $r_s$ from that point.

When $r_0 \ne 0$ one can repeat the above analysis. Now, one starts at
an $r > r_0$, and where $\frac{r_0}{r} < \frac{\Lambda r^2}{6}$. Then,
the analysis proceeds as before. If $\Lambda$ is positive (negative),
then the function $f$ will be decreasing (increasing), as $r$ increases
beyond $(\frac{6 r_0}{|\Lambda|})^{\frac{1}{3}}$. One then considers
cases (A), (B), and (C) as before, arriving at the same conclusion. This
is also physically reasonable since the cosmological constant can be
thought of as vacuum energy density and, as $r$ increases, the vacuum
energy overwhelms any non zero mass of a star, which is proportional to
$r_0$.

Thus, when the cosmological constant $\Lambda \ne 0$,
the static spherically symmetric gravitational field
of a point star in low energy string theory has a curvature
singularity, much worse than a naked singularity.

Requiring their absence
upto a distance $r_*$ then imposes a constraint on
$\Lambda$. If we take, somewhat arbitrarily, that the curvature becomes
unacceptably strong when $| \Lambda | r^2 \simeq 1$, then
$| \Lambda | r_*^2 < 1$, and we get the bound
\[
| \Lambda | < 10^{- 102} (\frac{r_*}{{\rm pc}})^{- 2}
\]
in natural units. Thus if $r_* \simeq 1 {\rm Mpc}$ then
$| \Lambda | < 10^{- 114}$, and if $r_*$ extends all the way upto the edge
of the universe ($10^{28} {\rm cm}$) then
$| \Lambda | < 10^{- 122}$ in natural units.

\vspace{4ex}

{\bf 5.}
We have analysed the static, spherically symmetric solutions to the \\
graviton-dilaton system, with or without electromagnetic couplings
and the cosmological constant. These solutions describe the
gravitational field of a point star. The main results are as follows.

1. For a charge neutral point star,
neither BD nor low energy string theory
predicts non trivial PPN parameters, $\beta$ and $\gamma$,
without introducing naked singularities.

2. With a cosmological constant $\Lambda$, coupled as in in low energy
string theory, the static spherically symmetric solutions are
likely to exist for BD type theories, with no naked singularities.
However, for low energy string theory, a non zero $\Lambda$ leads to
a curvature singularity, much worse than a naked singularity. Requiring
the absence of this singularity upto a distance $r_*$ implies a bound
$| \Lambda | < 10^{- 102} (\frac{r_*}{{\rm pc}})^{- 2}$ in natural units.
If $r_* \simeq 1 {\rm Mpc}$ then $| \Lambda | < 10^{- 114}$
and, if $r_* \simeq 10^{28} {\rm cm}$ then
$| \Lambda | < 10^{- 122}$ in natural units.

\vspace{2ex}

It is a pleasure to thank H. S. Mani and T. R. Seshadri for discussions
and, particularly, P. S. Joshi for communications regarding aspects of
singularity.

\end{document}